\documentclass[preprint]{aastex}
\usepackage{amssymb}
\usepackage{psfig}
 
\begin{document}

\title{First Results from MASIV: The Micro-Arcsecond Scintillation-Induced 
Variability Survey}

\author{J.~E.~J.~Lovell\altaffilmark{1} \email{Jim.Lovell@csiro.au}
D.~L.~Jauncey\altaffilmark{1} \email{David.Jauncey@csiro.au}
H.~E.~Bignall\altaffilmark{2,1,3} \email{bignall@jive.nl}
L.~Kedziora-Chudczer\altaffilmark{1} \email{Lucyna.Kedziora-Chudczer@csiro.au} 
J-P.~Macquart\altaffilmark{4} \email{jpm@astro.rug.nl}
B.~J.~Rickett\altaffilmark{5} \email{rickett@ece.ucsd.edu}
A.~K.~Tzioumis\altaffilmark{1} \email{Tasso.Tzioumis@csiro.au}}

\altaffiltext{1}{Australia Telescope National Facility, CSIRO, PO Box
76, Epping, NSW 1710, Australia}
\altaffiltext{2}{Department of Physics and Mathematical Physics, University of Adelaide, SA 5005, Australia}
\altaffiltext{3}{now at Joint Institute for VLBI in Europe, Postbus 2, 7990 AA, Dwingeloo,
The Netherlands}
\altaffiltext{4}{Kapteyn Astronomical Institute, University of
Groningen, Postbus 800 9700 AV, Groningen, The Netherlands}
\altaffiltext{5}{University of California San Diego, La Jolla,
 CA 92093}

\begin{abstract}
We are undertaking a large-scale, Micro-Arcsecond
Scintillation-Induced Variability (MASIV) survey of the northern sky,
$\delta > 0^{\circ}$, at 4.9~GHz with the VLA. Our objective is to
construct a sample of 100 to 150 scintillating extragalactic sources
with which to examine both the microarcsecond structure and the parent
populations of these sources, and to probe the turbulent interstellar
medium responsible for the scintillation. We report on our first epoch
of observations which revealed variability on timescales ranging from
hours to days in 85 of 710 compact flat-spectrum sources. The number
of highly variable sources, those with RMS flux density variations
greater than 4\% of the mean, increases with decreasing source flux
density but rapid, large amplitude variables such as J1819+3845 are
very rare.  When compared with a model for the scintillation due to
irregularities in a 500~pc thick electron layer, our preliminary
results indicate maximum brightness temperatures $\sim 10^{12}$~K,
similar to those obtained from VLBI surveys even though interstellar
scintillation is not subject to the same angular resolution limit.
\end{abstract}


\section{Introduction}

Considerable evidence has now accumulated to demonstrate that
interstellar scintillation (ISS) in the turbulent interstellar medium
(ISM) of our Galaxy, is the principal mechanism responsible for the
intra-day variability (IDV) seen in many flat-spectrum Active Galactic
Nuclei at centimeter wavelengths. Much of this evidence has come from
observations of the three fastest known IDV sources, B0405--385
\citep{ked97}, B1257--326 \citep{big2003} and J1819+3845
\citep{2000ApJ...529L..65D}.  The rapid variability in
these sources makes possible the detection and measurement of any
pattern delay at widely spaced radio telescopes, as might be expected
if ISS is the cause of the IDV.

Such a pattern time delay was first measured in B0405--385 between the
ATCA and the VLA at 5~GHz
\citep{jau2000} despite the unfavorable geometry for this measurement for such
a southern source.  This was soon followed by similar measurements
with J1819+3845 \citep{dtdb2002}, where the high northern declination
of the source and the northern location of both the VLA and the WSRT
made this a very elegant experiment. Most recently, a pattern time
delay has also been measured in B1257--326 between the ATCA and the
VLA at 4.9 and 8.5~GHz (Bignall 2003; Jauncey et al. 2003; Bignall et al., in preparation).

The presence of such a pattern time delay implies that the
turbulent structures in the ISM are moving at a speed close
to the 30~$\mbox{km s}^{-1}$ orbital speed of the Earth. For part of the
year the Earth and the ISM move together, the relative speed is low and the
scintillation pattern moves across the observer slowly.  Six months later
the Earth moves in the opposite direction so the relative speed is high and
the scintillation pattern moves more rapidly.

Such an ``annual cycle'' has now been reported in three sources,
J1819+3845 \citep{dtdb2001}, B0917+624 \citep{ric2001,jm2001} and
B1257-326 \citep{big2003}, and evidence is accumulating for its
presence in several more sources.  Moreover, an annual cycle has
been detected in B1257--326 not only in the characteristic time
scale, $T_{\rm char}$, but also in the time delay between the
scintillation patterns at 8.6 and 4.8~GHz. Bignall et al (2003)
suggest that this may be due to an offset between the central
components of the source at each frequency, as might be expected if
the source were jet-like on a microarcsecond scale, and optically
thick between 4.8 and 8.6~GHz. This second annual cycle occurs as the
ISM passes first across the 8.6~GHz then across the 4.8~GHz component,
with the apparent displacement on the sky due to the displacement of
the $\tau=1$ surfaces along the jet. If the position of the surfaces
scales as $\nu^{-1}$ along the jet \citep{1979ApJ...232...34B}, this
implies a displacement of 0.1~pc \citep{big2003}.

The above observations establish unequivocally ISS in the
local Galaxy as the principal cause of the IDV seen in these
sources. Moreover, the long time for which IDV has been seen
in some of these sources, more than a decade for B0917+624
\citep{fuh02} and more than 5 years for both B1257-326
and J1819+3845, suggests that such scintillating
components are relatively long lived despite their small
physical sizes.

A source must be small in order to scintillate. In the weak
scattering case, most commonly encountered at frequencies
$\gtrsim 5$~GHz, the source angular size, $\theta_{S}$, must
be comparable to the angular size of the first Fresnel zone
\citep[e.g.][]{narayan92}. This implies an angular size
\[\theta_S \lesssim \sqrt{\lambda/2 \pi D}\]
 which is at most tens of microarcseconds for screen
 distances, $D$, of tens to hundreds of parsecs. 

The limiting angular size for scintillation varies inversely with
the square root of the screen distance, and thus the brightness
temperatures inferred from ISS scale with the screen
distance. Reliable screen distances and thus brightness temperatures
have been estimated for five scintillators. Rickett et al. (1995)
found $6 \times 10^{12}$~K for B0917+624 for scattering at 200~pc. For
the three fast scintillators, B0405$-$385, B1257$-$326 and J1819+3845,
recent careful analyses has found very nearby screens with typical
distances of 10 to 30~pc
\citep{ric02,big2003,2000ApJ...529L..65D}. Such nearby screens yield
brightness temperatures of $2 \times 10^{13}$~K, $4 \times 10^{12}$~K
and $5 \times 10^{12}$~K respectively. For B1519$-$273 a lower limit
to the screen distance of 390~pc has been found, based on the measured
1.6~GHz angular size limit. This yields a brightness temperature lower
limit of $5 \times 10^{13}$~K, or possibly as high as $6 \times
10^{14}$~K \citep{mac2000}. These results imply large Doppler factors for these sources,
at least as high as several hundred \citep{rea94}, significantly
higher than those seen in existing VLBI surveys \citep{kel00,mar00}.

\section{The need for a large survey}

The realization that ISS is the principal physical process responsible
for IDV at centimetre wavelengths suggests that a large-scale
scintillation survey may address many questions regarding both source
microarcsecond structures and properties of the ISM not otherwise
accessible. To date, IDV surveys have been relatively small,
restricted to of order 100 of the strongest flat-spectrum sources
\citep{hee84,qui89,ked2001}. These surveys have shown that up to
$\sim$20\% of the strong sources scintillate, but the small size of these surveys has
yielded samples of no more than 20 scintillators, insufficient for
reliable statistical investigations.

Therefore we have undertaken a large scale 4.9~GHz Micro-Arcsecond
Scintillation-Induced Variability (MASIV) survey with the VLA. The aim
is to construct a large, statistically significant sample of 100 to
150 scintillating sources. 4.9~GHz was chosen as the observing
frequency as it is in the weak scattering regime where the variations
are rapid, but is close to the transition frequency where the largest
amplitude scintillations and shortest timescales are seen over most of the sky
\citep{wal98}. 

In order to understand the sources themselves, it is also necessary to
understand the effects of the ISM, since the two are intimately linked
via the scintillation process.  Assuming that the extragalactic
sources are distributed uniformly over the sky, the observed
distribution of 100 or more scintillators may shed light on the
distribution of scattering material throughout the northern Galaxy. In
particular, the presence of new fast scintillators can reveal the
presence of nearby scattering material. Furthermore, determination of
the annual cycle distribution will lead to the overall distribution of
the velocity of the ISM.

For the sources themselves, the survey results lead directly to an
understanding of the dependence on scintillation on flux density,
spectral index, the nature of optical counterpart, redshift, luminosity,
milliarcsecond structure and evolution.  Most importantly, the survey allows us to
better characterize the scattering process responsible and hence
constrain the source brightness temperatures.

To make sure that the survey is unbiased with regard to characteristic
scintillation time scales, all sources were observed over three
well-spaced epochs during our observations, so that sources in the ``slow''
part of their annual cycle would not be missed. Three well-spaced
epochs also allows us to better find those sources like B0405$-$385
whose scintillation outbursts are episodic \citep{ked97}. Each session
extended over three or more days to prevent missing the slower
scintillators. Moreover, the 2 hour minimum observing interval gave
sufficient coverage each day to reliably recognize any variability.

We expect that any new fast scintillators found in the survey will
reveal the presence of nearby regions of interstellar turbulence in
the Galaxy.  Both of the long-lived fast scintillators are among the
weakest known, so the survey included a
selection of both strong, $\sim$1~Jy, and weak, $\sim0.1$~ Jy,
sources. Moreover, as the VLA was in a variety of configurations
during the year, we selected only sources that would appear point-like to the VLA at all resolutions. We selected compact flat-spectrum
sources as they are expected to possess a compact nucleus. The sources
were selected from the JVAS
\citep{1992MNRAS.254..655P,bro98,1998MNRAS.300..790W} and
CLASS \citep{1995ApJ...447L...5M} catalogs of sources unresolved at 8.5~GHz
with the VLA. From the JVAS catalog we selected all sources with
greater than 95\% of their flux density in an unresolved component and
from the CLASS survey we selected all sources with modeled source
sizes less than 50 mas. Sources were rejected if the catalogs
indicated the presence of confusing sources in the field.  We
cross-correlated these sources with the NVSS catalog to obtain
spectral indices of all sources and thus a flat-spectrum sample. We
chose a spectral-index lower-limit of $-0.3$ ($S \propto \nu^\alpha$)
which provided a sample of 1871 sources stronger than 100~mJy at
8.5~GHz. 

To reduce this to a manageable quantity for $\sim$2 hourly
flux density monitoring, the sample was further reduced into strong
and weak sub-samples of approximately 300 sources each. The weak sample
consists of sources with 8.5~GHz flux densities between 105 and
130~mJy and the strong sample contains sources stronger than
600~mJy. We also selected a sub-sample of sources between 130 and
600~mJy with $4.5~\mbox{h}
\leq \mbox{RA} < 7.5~\mbox{h}$ or $16.5~\mbox{h}
\leq \mbox{RA} < 19.5~\mbox{h}$ and $\delta \geq 35^{\circ}$.
These two ``windows'' coincide with regions of the sky where the
predicted ratio between the fastest and slowest variability timescale
over the course of a year are at their minimum ($\mbox{RA} \approx
6~\mbox{h}$) and maximum ($\mbox{RA} \approx 18~\mbox{h}$) for ISS-induced
variability in a plasma moving with the LSR \citep{2002PASA...19..100R}, and
thus let us probe a full range of flux densities for these two
regions. The declination limit was chosen to avoid excessive azimuth
slew times at the VLA for sources near transit. The total number of
sources in our sample is 710.

\section{Observations}

Our first epoch of observations spanned 72 hours from 2002 January 19
to 2002 January 22 during reconfiguration from D to A array. At the
start of our observations all antennas had been moved into A-array
except for two on the south spur of the array, separated by 40~m. The
observations were conducted in a five-subarray mode. The combined
strong and weak samples were divided into four declination bands and
one subarray was assigned to each. The fifth subarray was assigned to
the intermediate flux-density sources in the minimum and maximum
annual modulation ratio windows defined above.

Each subarray was scheduled so that every source was observed for one
minute every $\sim$2 h while it was above an elevation of
$15^\circ$. Antennas were assigned to subarrays to form long baselines
to reduce possible contributions from any extended structure. Shadowing was avoided
for the two antennas in D-array by assigning the northern-most antenna
to the northern-most declination band and the southern-most to the
southern declination band. For flux density calibration, each subarray
observed B1328+307 (3C286) and J2355+4950 every $\sim$~2
hours. B1328+307 is the primary flux density calibrator for the VLA and
J2355+4950 is a GPS source, not likely to vary over short timescales,
and is monitored regularly at the VLA as part of a calibrator
monitoring program.

The data from our observations were loaded into AIPS in real-time
using the on-line FILLM task, thus providing quick access to the
data. After the first 24~h we performed a rudimentary amplitude
calibration and began searching for evidence of short timescale
variability with peak-to-peak amplitudes greater than $\sim$10\%, well above our estimated
systematic errors of a few percent. Once these sources were identified we
scheduled additional scans on them in the last 24 h of our
observations to improve the sampling and hence our estimation of the
variability timescale.

Following the observations we applied a more rigorous calibration of
both total intensity and linear polarization data. This required
particular care as many antennas had recently been moved and their
pointing calibration observations were not complete. The residual
pointing errors may depend on azimuth and
elevation so we chose several bright, non-variable sources in each
subarray at a range of right ascensions as gain calibrators for
surrounding sources. Precautions were taken to ensure that the
calibrators themselves were not variable: if a given calibrator caused
the majority of sources against which it was applied to vary, then
another calibrator was chosen.

 As the subarray beam pattern rotates on the sky, the presence of
 arcsecond scale structure adds alternatively in-phase and
 out-of-phase with the central unresolved ``core'' creating the
 appearance of variability. However, it then has the same variability
 character on each of the three days observations, and such a
 repetitive pattern is relatively easily seen in the data. Moreover,
 imaging the three day data set quickly resolves any
 ambiguities. Snapshot images were therefore made of each source
 showing evidence of variability to guard against such spurious
 detection of IDV. This is most important when the VLA was in its more
 compact configurations.

\section{Results}
\label{sec:four}

\begin{figure}[htbp]
  \centerline{\psfig{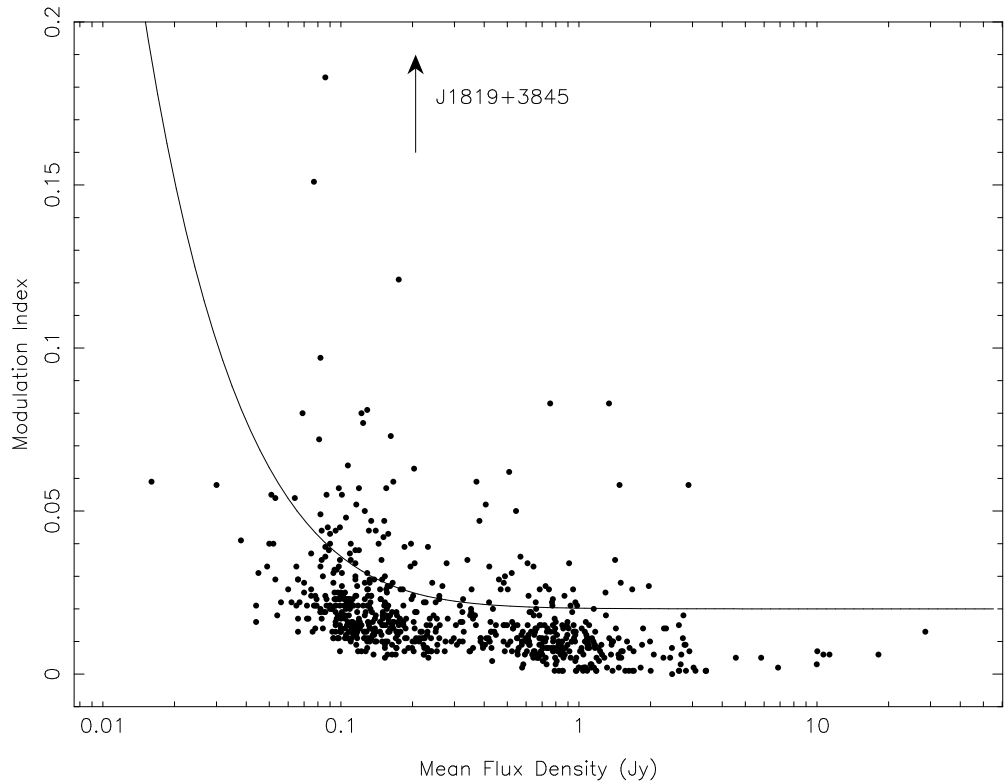}}
  \caption{Variability index $(\mbox{rms}/\bar{S})$ as a function of
mean flux density. We classify any source above the solid line, as defined in Section~\ref{sec:four}, as variable. }
  \label{fig:sig_vs_sbar}
\end{figure}

In order to separate variable from non-variable sources it is
essential to  determine reliably the uncertainties in the VLA flux
density measurements.  As a first step we plotted, in Figure~\ref{fig:sig_vs_sbar}, the
modulation index (RMS divided by the mean flux density), versus mean
flux density, for each of the 710 sources in the survey.  Figure 1
shows clearly the flux density groupings of the strong and weak
samples, and that the great majority of the sources have average RMS
errors well below 5\%.

Uncertainties in flux density measurements are made up of two
components, one, $s$ Jy, due to noise and confusion, that is
independent of flux density, and the other, $p$ \%, that is
proportional to flux density and due primarily to pointing offsets.
The dominant uncertainty among the strong sources is $p$ while $s$ is
the main contributor among weak sources.  Examination of Figure~\ref{fig:sig_vs_sbar}
indicates that $s$ is close to 1.5 mJy and $p$ is close to 1\%. As a
first step, we then selected for a more detailed examination, those
sources with

\[ \mbox{RMS} \geq \sqrt{(2s)^2 + \left( \bar{S} \frac{2p}{100}\right)^2}\,\,\, \mbox{Jy}\]

where $\bar{S}$ is the mean flux density at 4.9~GHz.
This corresponds to a simple $\chi$-squared test to identify those sources
whose scatter is well in excess of the overall population errors; that is
those that are likely variables. This procedure identified 99 sources.
While this is not the most effective way of finding the variable sources,
it serves as an effective starting point to determine the errors for the
survey as a whole.

\begin{figure}[htbp]
  \begin{center}
    \leavevmode
        \psfig{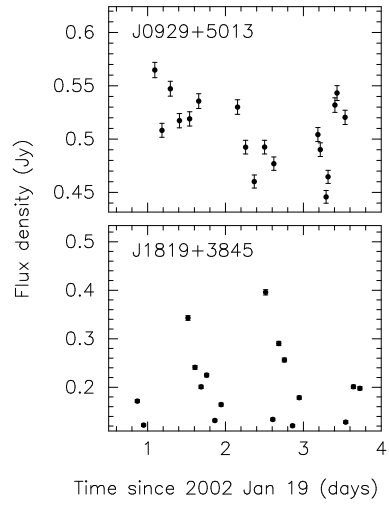}
  \end{center}
  \caption{Lightcurves for two of the most rapid variables found in
the survey: J0929$+$5013 and the previously known IDV J1819$+$3845.}
  \label{fig:lc1}
\end{figure}

\begin{figure}[htbp]
  \begin{center}
    \leavevmode
        \psfig{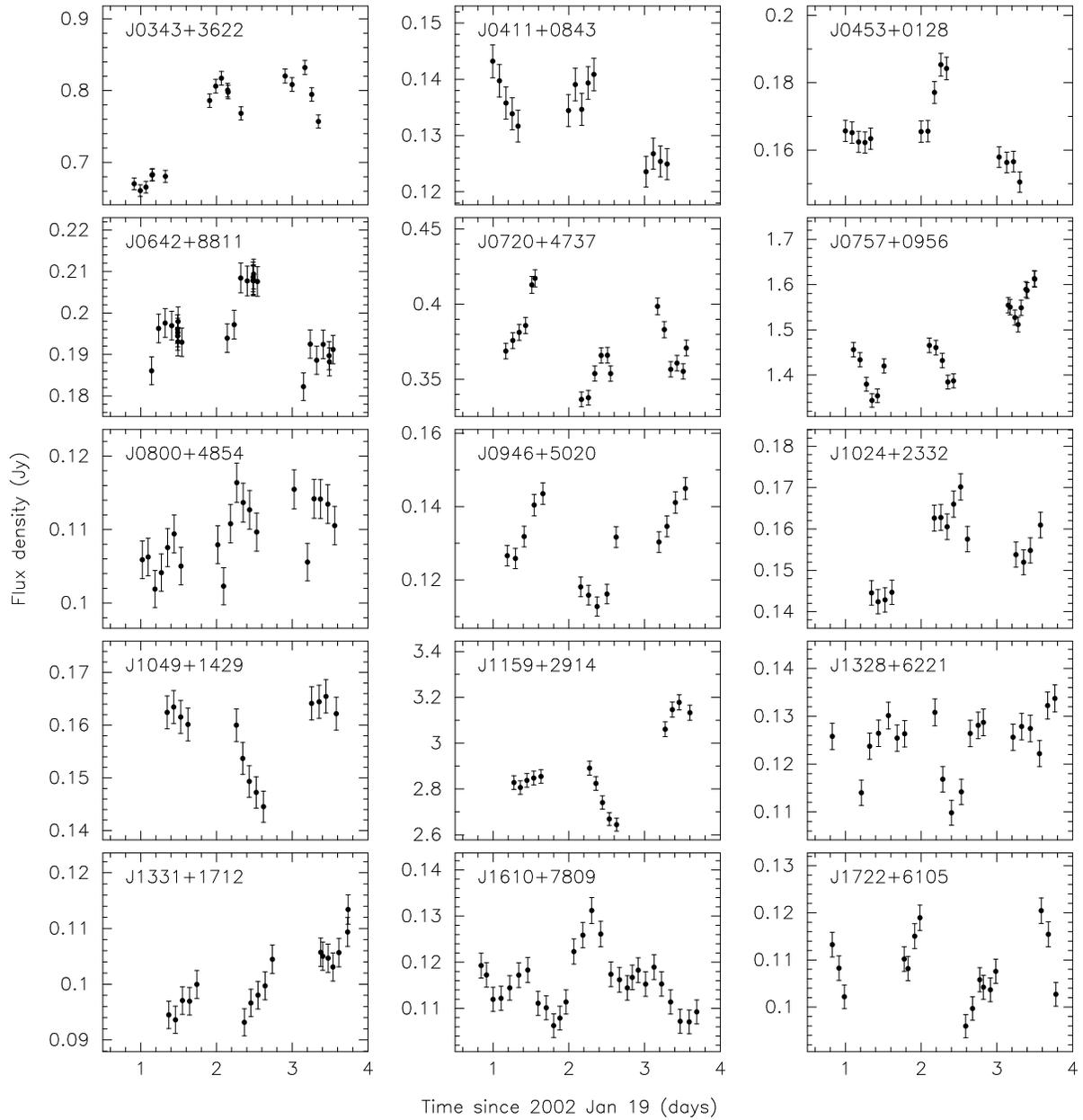}
  \end{center}
  \caption{Lightcurves of variable sources with timescales ranging
from a few hours to approximately one day.}
  \label{fig:lc2}
\end{figure}

\begin{figure}[htbp]
        \psfig{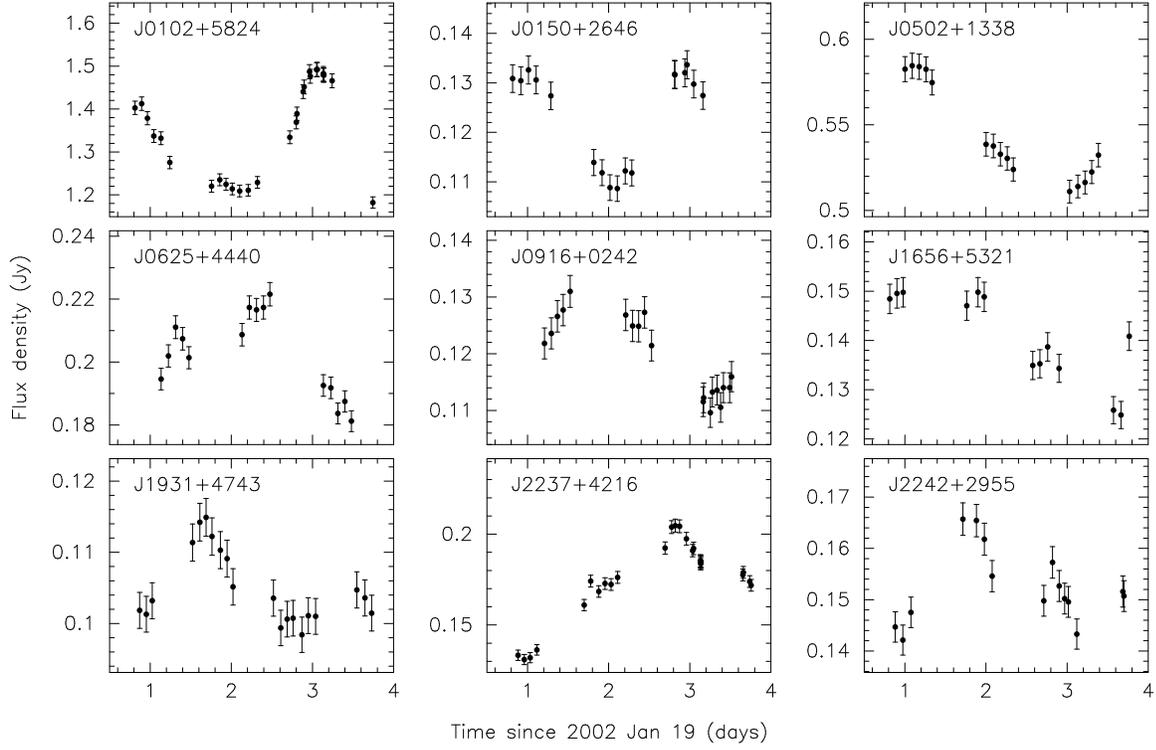}
  \caption{Lightcurves of sources varying on timescales of $\sim$1 to
$\sim$2 days.}
  \label{fig:lc3}
\end{figure}

\begin{figure}[htbp]
        \psfig{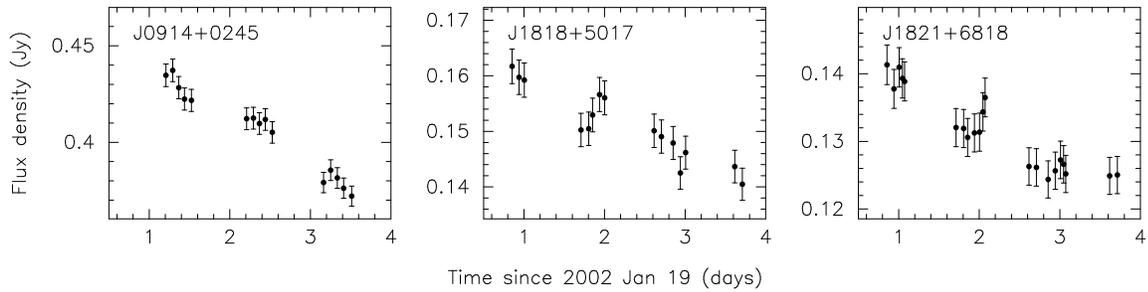}
  \caption{Lightcurves of sources varying on timescales longer than the period of the observations.}
  \label{fig:lc4}
\end{figure}

Data collected for each of these 99 sources were then examined in
detail. In many instances not only were there obvious strong flux
density changes, but the pattern of the variations are remarkably
similar to those seen in many of the known IDV sources
\citep{ked2001}. To give a flavor of the range of variations seen, and
to provide an initial catalog of sources from which others may wish to
conduct more detailed investigations, we list in
Table~\ref{tab:sources} all sources with a mean flux density greater
than 100~mJy and modulation indices of 0.04 or more, making a total of
29 sources. Light curves for these sources are also shown in
Figures~\ref{fig:lc1}-\ref{fig:lc4} and we now discuss these sources
individually from the shortest to the longest variability timescales.

Variability was detected on a full range of timescales from less than
two hours to greater than three days.(We define timescale here as the mean
time between a peak and a minimum in a light-curve). In
Figure~\ref{fig:lc1} we show two of the most rapid variables detected:
J0929$+$5013 and J1819$+$3845. Both are clearly under-sampled in our
survey observations but are detected as variable
nonetheless. 

J0929$+$5013 is a BL Lac object
\citep{1996A&A...309..419N}. Like most strong ($S_{\rm 4.9\, GHz} > 0.3$~Jy) sources
listed in Table~\ref{tab:sources}, it has been observed at
milliarcsecond-scale resolution and is compact
\citep{2002ApJS..141...13B}.  

J1819$+$3845 was serendipitously
discovered to be an IDV 3 years ago \citep{2000ApJ...529L..65D}, and exhibits the
most rapid flux density variations known for any extragalactic
source. The presence of an annual cycle in the time scale for
variability \citep{dtdb2001}, plus the measurement of a varying time
delay between the IDV pattern arrival times at the VLA and WSRT
\citep{dtdb2002} establishes unequivocally that ISS is the mechanism
responsible for the rapid flux density variability in this source. It
is remarkable how readily the IDV can be seen in this source, even
sampling more slowly than the characteristic time scale for
variability.  Clearly rapid variables like J1819$+$3845 would not
have been missed in our survey.

Lightcurves for sources that vary on  longer timescales which
are well sampled by our observations are shown in
Figure~\ref{fig:lc2}. Of particular note is J1159+2914 (B1156+295), a
quasar at $z=0.729$ which has the highest RMS flux density observed in
our survey at 167~mJy. This object has also been identified as a
source of gamma-rays \citep{1997ApJ...490..116M} and is highly
variable
\citep[e.g.][]{1995IAUC.6168....1W,1994A&AS..106..361X,1997A&AS..121..119V}.
VLBI observations reveal a high brightness temperature nucleus
\citep{1998Sci...281.1825H} and apparent superluminal jet speeds of up to 
$9$~c ($H_0 = 65 \,\mbox{km
s}^{-1}\mbox{Mpc}^{-1}$, $q_0 =0.1$) have been claimed
\citep{1997ApJ...485L..61P,2001ApJS..134..181J}. 

J0757$+$0956 is a BL Lac \citep{2000A&A...357...91F}.  It shows variability on two different
timescales with quite rapid changes on a $\sim 0.3$~day timescale and
a slow overall rise in flux density over three days. This is perhaps
indicative of scintillation of two sub-components in the source, one
smaller than the other, or perhaps scintillation in a small component
superimposed on slower intrinsic variability. Similar multi-timescale
behavior is also seen in J0343+3622, J0453+0128 and J0800+4854.
The continued monitoring from the remaining two epochs of our VLA program
will help distinguish intrinsic from scintillation-induced variability
through a search for changes in timescale.

J1049$+$1429 is one of the more unusual variables and highlights the
benefit of a 72~h observation. Very little variability is seen on the
first and third days but on the second day we observe a $\sim 10$\%
drop in flux density. This is consistent with variations on a
characteristic time of about 1~day, in which the flux density on the
first and third days happened to be nearly equal.

J1331+1712 also shows some evidence of variability on two different
timescales. There is an overall increase in flux density seen over the
three days together with what appears to be short-term variability on
the timescale of about 1 day. In this case however the apparent
short-term variability is due to resolution effects from a previously
undetected extended $\sim$1 arcsec jet which contributes 15~mJy to the
total flux density. This source was therefore not included in later
MASIV survey observations.

Sources with intermediate variability timescales of 1 to 2 days are
shown in Figure~\ref{fig:lc3} and demonstrate the value of a 72~h
observation in detecting slow variables. It is clear from the
light-curves that variability in many such sources could have gone
undetected with shorter monitoring periods of 2 days or less.
J0102$+$5824 (B0059$+$581) shows variations on a timescale of $\sim$~1
day. This source has been the subject of daily monitoring as part of
the Keystone Project \citep{koy2001} and shows a clear annual cycle
in its variability at 2.3~GHz (Jauncey et al. in preparation) with
timescales ranging from 10 to $> 100$~days. As expected for ISS, the
variations are faster at 4.9~GHz and we expect to see similar changes
in timescale over the course of our VLA observations.

J2237+4216, an object in our sample of weak sources, has a
very high modulation index, displaying a doubling in flux
density over 2 days. Little is known about this source at
present but we are pursuing optical identification and VLBI
imaging for this source and all other variables yet to be identified or imaged.

Varying on the longest detectable timescales are J0914+0245,
J1818+5017 and J1821+6818 (Figure~\ref{fig:lc4}).  They show a
monotonic change in flux density and therefore only lower limits on
their variability timescales can be estimated. If these slower changes
over the three days are intrinsic in origin, then the implied
variability brightness temperatures \citep{1995ARA&A..33..163W} are greater than $10^{16}$~K. This is well in excess of any scintillation
brightness temperatures seen to date, and suggests that the IDV may
have a scintillation, rather than intrinsic, origin. It will be of
considerable importance to see if these sources exhibit an annual
cycle which is the hallmark of ISS.

The above sources establish the ability of the survey to detect strong
IDV, but it remains important to determine the limits to this
ability. We are addressing this question through a detailed individual
analysis of the 99 sources identified as variable. While the RMS
variations in the light curves give a good indication of the flux
density measurement errors for the full sample, the variability
selection criterion based on the RMS alone cannot be applied blindly
to individual sources. Our analysis found that in 14 of these 99
sources, the observed flux density changes were due to other causes
than real variability: in particular, the presence of weak structure
or weak, nearby confusing sources.

\subsection{Source Properties and Statistics}

 Here we describe some general properties of the sources observed in
the first epoch of
 the survey. A more detailed description of all variable sources will
 be left to future papers.

Of the 710 sources observed in the first epoch, a total of 85 sources (12\%) are
classified as variable by our RMS selection criterion.  From
Figure~\ref{fig:sig_vs_sbar} it appears that there are more variable
sources with high modulation indices in the weaker sample than the
strong sample. To test this we divided the sources into strong and
weak samples based on their 4.9~GHz flux density measured in our first
epoch of observations. The weak sample was defined as all sources with
a mean flux density less than 0.3~Jy with the remainder placed into
the strong sample. We chose to use flux densities at 4.9~GHz instead
of the 8.5~GHz JVAS and CLASS flux densities because intrinsic,
long-term variability tends to be less at lower frequencies.
We then selected a sample of high modulation index variables which we
define as having a RMS/mean $\geq 0.04$. This level is well above the
combined systematic (pointing) and sensitivity limits of our
observations. A total of 13 sources had mean flux densities less than
60~mJy and were excluded from this analysis as our ability to detect
4\% variability begins to be affected by the errors below this
level. We found that of the 363 weak sources, 33 were highly variable
while only 8 of the 320 strong sources were highly variable. To test
this apparent difference in populations we constructed a simple
chi-squared contingency test with a null hypothesis that all sources
with a fractional variation $\geq 0.04$ come from the same population
and found that the probability this is true is less than
0.5\%. Therefore we detect significantly more variables with high
modulation indices in the weaker than the stronger source population.

\section{Discussion}

It is remarkable that of the 710 sources observed
there are none that vary to the same degree as J1819$+$3845, a source
whose intra-day variability was discovered serendipitously,
and whose variability is caused by enhanced scattering
within 20-40~pc of the Earth \citep{2000ApJ...529L..65D}. 

Of the ten sources in Table~\ref{tab:sources} that already have
optical spectroscopy, five are BL Lac objects and five are
quasars. Such a high fraction of BL Lacs among the scintillators is in
general agreement with the optical identification content of existing
surveys \citep{qui2000,ked2001}.

We find there are more highly variable sources in the weaker part of our sample.
There are at least two possible explanations for this. Firstly, as
stated earlier, the micro-arcsecond scintillating components may
simply be brightness temperature limited,  in which case the weaker
sources may simply be smaller  and hence are more likely
to scintillate than the stronger sources.  Alternatively, this effect
may be the result of sampling different source populations where
weaker sources are more ``core dominated'', or rather less
milli-arcsecond ``jet dominated''.  To help distinguish between these
two possibilities we are imaging the 55 weakest scintillating sources
with the VLBA to determine their morphology and place limits on their
brightness temperatures. 
Follow-up
observations one year later will determine the fraction of these sources
that show proper motions.

We have calculated the scintillation index from a simple model based
on weak scintillation theory using the Taylor and Cordes (1993) [TC93]
model for the interstellar electron density and a brightness-limited
model for each source.  We tested two source models, in which half of
the total flux density is in a Gaussian component with peak brightness
of $10^{11}$K or $10^{12}$K, with the remaining flux density in
regions too large to scintillate ($>0.1$ mas).  For the $10^{11}$K
model the predicted scintillation indices were barely detectable ($\la
0.02$), but for $10^{12}$K they were in the range 0.02 - 0.25 and are
similar to those seen in our scintillators.  From this we tentatively
conclude that the 85 scintillators have peak brightnesses in the
neighborhood of $10^{12}$K and the remaining sources have most of
their flux density from brightness temperatures $\la 10^{11}$K .  Any
decrease in the scattering distance below the 500~pc typical of the
TC93 model reduces the implied brightness.  Values above $10^{12}$K
are only possible from sources with less than half of their flux
density in the compact component. This brightness temperature limit is
similar to those obtained from VLBI observations even though ISS is
not subject to the same angular resolution limits.

The remaining epochs of our initial survey have now been observed. In
future papers we will present a complete catalog of variable sources
and a detailed examination of their properties. These investigations
will include a comparison of changes in variability timescale with
those expected due to the changing velocity of the earth relative to
the LSR. Those sources that do not show these expected ISM-induced
changes will be of particular interest as they may be revealing
peculiar screen velocities, intrinsic flux density variability or
changes in source structure that induce episodic variability.

It would seem that the boundary between intrinsic and
scintillation-induced variability is not yet fully explored as
demonstrated by our detection of sources that change on timescales
greater than $\sim$3 days. For these sources it may be necessary to
conduct daily observations over the course of at least a year to
determine if annual cycles exist.

We are already undertaking follow-up observations of the newly
detected IDVs including more intensive monitoring of the rapid
variable sources with WSRT and VLBI snapshot imaging of the IDVs with
previously unknown milliarcsecond-scale structures. We are also
undertaking low-frequency VLBI observations of some of the brighter
IDVs to search for evidence of scatter-broadening which will provide
additional constraints on the size of the scattering disk.

With only a limited amount of VLBI data currently available on the
variable sources it is not yet possible to draw conclusions or
identify obvious trends on the relationship between milliarcsecond and
microarcsecond-scale structures. The new VLBI observations we are
undertaking will allow us to address this issue.

Several of the new IDVs are regularly observed as part of geodesy and
astrometry programs and some may be target phase reference sources for
programs such as VERA \citep{2000SPIE.4015..624H}. A number of the
weak scintillators may be amongst the most compact sources to be found
at VLBI resolution and hence may form part of the astrometric radio
reference frame in the future. It is important to understand the
impact of scintillation. As the scintillating component in the source
changes intensity as seen by a VLBI array, the observed phase center
of the source will also move on the sky. The amount of movement will
depend on the location and flux density contribution of the
scintillating component with respect to the rest of the source.

\acknowledgments

We are extremely grateful for the technical support provided by NRAO
staff at Socorro, in particular we would like to thank Ken Sowinski,
Miller Goss, Mark Claussen and Jim Ulvestad. We would also like to
thank Neal Jackson for providing the catalog of CLASS sources from
which part of our sample was drawn. The National Radio Astronomy
Observatory is a facility of the National Science Foundation operated
under cooperative agreement by Associated Universities, Inc. HEB
acknowledges the support of a Faculty of Science Scholarship from the
University of Adelaide. BJR
thanks the US-NSF for funding under grant AST 9988398.

\begin{deluxetable}{rlccccl}
\tablewidth{0pt}
\tablecaption{Variable sources detected in the first epoch of
MASIV survey observations with mean 4.9~GHz flux density greater than
100~mJy and modulation indices of 0.04 or more.\label{tab:sources}}
\tablehead{
\colhead{J2000 Name} &
\colhead{B1950 Name} &
\colhead{$\bar{S}_{4.9}$ (Jy)} &
\colhead{$100\sigma/\bar{S}_{4.9}$} &
\colhead{ID} &
\colhead{z} &
\colhead{Ref} }
\startdata
 JVAS J0102$+$5824 & B0059$+$581  &   1.34   &   8.3  & \nodata & \nodata & \nodata \\
CLASS J0150$+$2646 & B0147$+$2631 &   0.12   &   7.7  & \nodata & \nodata & \nodata \\
 JVAS J0343$+$3622 & B0340$+$3612 &   0.76   &   8.3  &  Q  & 1.484   & 1   \\
 JVAS J0411$+$0843 & B0408$+$0835 &   0.13   &   4.7  & \nodata & \nodata & \nodata \\
 JVAS J0453$+$0128 & B0450$+$0123 &   0.17   &   5.9  & \nodata & \nodata & \nodata \\
 JVAS J0502$+$1338 & B0459$+$135  &   0.54   &   5.0  &  BL  & \nodata &  2  \\
 JVAS J0625$+$4440 & B0621$+$4441 &   0.20   &   6.3  &  BL  & \nodata &  1  \\
 JVAS J0642$+$8811 & B0604$+$8813 &   0.20   &   4.0  & \nodata & \nodata & \nodata \\
 JVAS J0720$+$4737 & B0716$+$4743 &   0.37   &   5.9  & \nodata & \nodata & \nodata \\
 JVAS J0757$+$0956 & B0754$+$100  &   1.48   &   5.8  &  BL  & \nodata &  3  \\
CLASS J0800$+$4854 & B0756$+$4902 &   0.11   &   4.0  & \nodata & \nodata & \nodata \\
 JVAS J0914$+$0245 & B0912$+$029  &   0.41   &   5.2  &  Q  &  0.427  & 4   \\
CLASS J0916$+$0242 & B0914$+$0255 &   0.12   &   5.7  & \nodata & \nodata & \nodata \\
 JVAS J0929$+$5013 & B0925$+$5026 &   0.51   &   6.2  &  BL  & \nodata &  5  \\
CLASS J0946$+$5020 & B0942$+$5034 &   0.13   &   8.1  & \nodata & \nodata & \nodata \\
CLASS J1024$+$2332 & B1022$+$2347 &   0.16   &   5.7  &  BL  & \nodata &  6  \\
CLASS J1049$+$1429 & B1047$+$1445 &   0.16   &   4.3  & \nodata & \nodata & \nodata \\
 JVAS J1159$+$2914 & B1156$+$295  &   2.89   &   5.8  &  Q  & 0.729   &  7  \\
CLASS J1328$+$6221 & B1326$+$6237 &   0.13   &   5.0  & \nodata & \nodata & \nodata \\
CLASS J1331$+$1712 & B1329$+$1727 &   0.10   &   5.5  & \nodata & \nodata & \nodata \\
 JVAS J1610$+$7809 & B1612$+$7817 &   0.12   &   5.2  & \nodata & \nodata & \nodata \\
 JVAS J1656$+$5321 & B1655$+$5326 &   0.14   &   4.4  & Q   & 1.553   &  8  \\
 JVAS J1722$+$6105 & B1722$+$6108 &   0.11   &   6.4  & \nodata & \nodata & \nodata \\
CLASS J1818$+$5017 & B1817$+$5015 &   0.15   &   4.2  & \nodata & \nodata & \nodata \\
CLASS J1819$+$3845 & B1817$+$3843 &   0.21   &  37.5  & Q   & 0.54   &   9 \\
 JVAS J1821$+$6818 & B1822$+$6817 &   0.13   &   4.4  & \nodata & \nodata & \nodata \\
CLASS J1931$+$4743 & B1929$+$4737 &   0.11   &   4.8  & \nodata & \nodata & \nodata \\
CLASS J2237$+$4216 & B2234$+$4201 &   0.18   &  12.1  & \nodata & \nodata & \nodata \\
CLASS J2242$+$2955 & B2239$+$2939 &   0.15   &   4.7  & \nodata & \nodata & \nodata 
\enddata
\tablecomments{Identifications (Q = quasar, BL = Bl Lac) and
redshifts are given where available. Reference are
(1) \citet{1995AJ....109.1983V}; (2) \citet{1998AJ....115.1253P};
(3) \citet{2000A&A...357...91F}; (4) \citet{1997MNRAS.284...85D};
(5) \citet{1996A&A...309..419N}; (6) \citet{2000A&A...356..445B};
(7) \citet{1994ApJS...93....1A}; (8) \citet{1998ApJ...494...47F};
(9) \citet{2000ApJ...529L..65D} }

\end{deluxetable}

\end{document}